\documentclass[aps,twocolumn,showpacs]{revtex4}
\usepackage{graphicx}
\usepackage{amsmath}
\usepackage{amssymb}

\begin{document}

\title{Integrability-breaking phase transitions in stadium-like billiards}

\author{Anne K\'etri P. da Fonseca, Edson D.\ Leonel}

\affiliation{Departamento de F\'isica, Unesp - Universidade Estadual Paulista - 
Av.24A. 1515, 13506-700, Rio Claro, SP, Brazil}

\date{\today} \widetext

%\pacs{05.45.-a, 05.45.Pq, 05.45.Tp}

\begin{abstract}
We investigate integrability-breaking transitions in two classes of stadium-like billiards with parabolic boundaries. While focusing boundaries generate a mixed phase space in which regular islands coexist with a chaotic sea, dispersing boundaries produce a fully chaotic phase space for any finite boundary deformation. By analyzing the scaling behavior of the roughness $\omega$, we identify two qualitatively distinct transitions: a continuous transition for the focusing geometry and a first-order transition for the dispersing one. We determine the corresponding critical exponents and establish the associated scaling laws. For the continuous transition, we further provide a complete characterization within the framework of critical phenomena by identifying the broken symmetry, the order parameter and its diverging susceptibility, the elementary excitations responsible for chaotic diffusion, and the topological defects governing transport. These results establish a statistical-mechanics framework for describing integrability-breaking transitions in Hamiltonian billiards and suggest that the concepts of critical phenomena naturally extend to deterministic nonlinear dynamical systems.
\end{abstract}

%\keywords{Suggested keywords}%Use showkeys class option if keyword
                              %display desired
\maketitle

%\tableofcontents

\section{Introduction}
\label{introduction}

Phase transitions provide one of the most successful frameworks for understanding the emergence of collective behavior in physical systems. Their characterization relies on a set of fundamental concepts, including broken symmetry, order parameters, elementary excitations, topological defects, and critical scaling. While these ideas were originally developed within statistical mechanics to describe equilibrium phenomena such as ferromagnetism \cite{khanna1991magnetic,grinstein1976ferromagnetic,gutzwiller1963effect}, superconductivity \cite{kiometzis1994critical,bianchi2002first,vojta2000quantum}, and Bose-Einstein condensation \cite{wouters2010superfluidity,berman2008bose,chen2000unusual}, they naturally raise an important question in nonlinear dynamics: can the transition from integrability to chaos be characterized within the same conceptual framework?

Integrability-breaking transitions arise naturally in Hamiltonian systems described by the generic Hamiltonian
$H(I_1,\theta_1,I_2,\theta_2)=H_0(I_1,I_2)+\epsilon H_1(I_1,\theta_1,I_2,\theta_2)$.
Representative examples include the standard mapping \cite{leonel2020characterization}, collision models such as the Fermi-Ulam accelerator \cite{11nova} and the bouncing ball model \cite{oliveira2013some}, and the twist mapping \cite{nontwist}. For $\epsilon=0$, the dynamics is fully integrable and the phase space is foliated by invariant curves supporting periodic or quasiperiodic motion. As soon as $\epsilon\neq0$, chaotic trajectories emerge. For sufficiently weak perturbations, regular islands coexist with chaotic regions, forming the well-known mixed phase space \cite{leonel2015dynamical}. Near this transition, scaling invariance frequently emerges and is characterized by a set of critical exponents \cite{leonel2005scaling}. The resulting scaling collapse provides compelling evidence of critical behavior \cite{pathria2011statistical}. Nevertheless, scaling alone does not establish a complete analogy with phase transitions, since the remaining ingredients of the statistical-mechanics framework are generally absent.

Billiard systems constitute an ideal setting for addressing this problem. They describe the motion of one or more non-interacting particles undergoing specular reflections inside a closed boundary \cite{chernov2006chaotic}. Since the dynamics are entirely determined by the boundary geometry, billiards provide a particularly clean framework in which integrability can be continuously broken through geometric deformations \cite{robnik1983classical}. Whereas the circular billiard is completely integrable, even simple boundary deformations may generate chaotic dynamics \cite{Bunimovich1979}.

\begin{figure}[h]
  \centering
    \centerline{\includegraphics[width=0.9\linewidth]{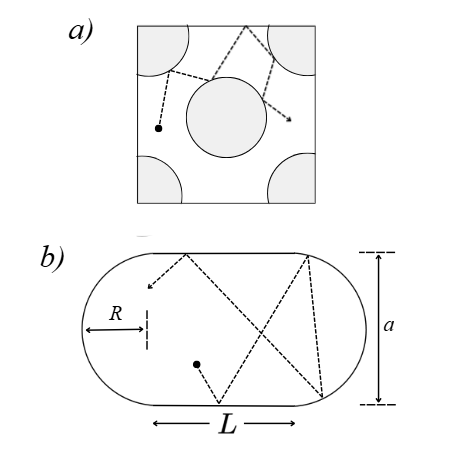}}
  \caption{Sketch of (a) a dispersing billiard, illustrating multiple scattering events analogous to the Lorentz gas, and (b) a Bunimovich stadium with its corresponding geometric parameters. In both panels, a representative particle trajectory is shown.}
  \label{Fig1}
\end{figure}

Depending on their geometry, billiards are commonly classified as dispersing, focusing, or neutral. Dispersing billiards, first introduced by Sinai \cite{Sinai70}, consist exclusively of neutral and dispersing boundary components. A paradigmatic example is the Lorentz gas (Fig.~\ref{Fig1}a), whose deterministic dynamics exhibit transport properties analogous to Brownian motion \cite{17}. Focusing boundaries, on the other hand, typically generate mixed phase spaces in which regular islands coexist with chaotic regions \cite{chernov2006chaotic}. The Bunimovich stadium, formed by two semicircles connected by parallel straight segments, is the best-known example of this class \cite{Bunimovich1979}. Replacing the semicircular arcs by parabolic boundaries \cite{loskutov2002} introduces a continuous geometric control parameter $b$, whose magnitude and sign determine both the strength and the nature of the nonlinear perturbation. While $b=0$ corresponds to a completely integrable rectangular billiard, positive and negative values generate dispersing and focusing geometries, respectively.

In this work, we investigate the transition from integrability to non-integrability in two families of stadium-like billiards generated by opposite curvatures of the parabolic boundaries. For dispersing boundaries ($b>0$), the phase space becomes fully chaotic for any finite perturbation. In contrast, focusing boundaries ($b<0$) produce a mixed phase space in which regular islands coexist with a chaotic sea before the onset of global chaos. We characterize both transitions through their scaling properties and critical exponents and, more importantly, establish a complete statistical-mechanics description of the continuous transition by identifying its broken symmetry, order parameter, elementary excitations, and topological defects. This framework not only reveals a previously unreported universality class for integrability-breaking transitions but also demonstrates that the conceptual language of critical phenomena naturally extends to deterministic Hamiltonian dynamics.

The remainder of this paper is organized as follows. Section \ref{xsec2} introduces the model and derives the mapping for stadium-like billiards with focusing and dispersing boundaries. Section \ref{xsec3} presents the scaling analysis and the corresponding critical exponents. In Section \ref{xsec4}, we establish the statistical-mechanics characterization of the transition by identifying its broken symmetry, order parameter, elementary excitations, and topological defects. Finally, Section \ref{xsec5} summarizes the main results and presents our conclusions.

\section{The model and the mapping}
\label{xsec2}

\begin{figure}[h]
\centering
    \centerline{\includegraphics[width=0.65\linewidth]{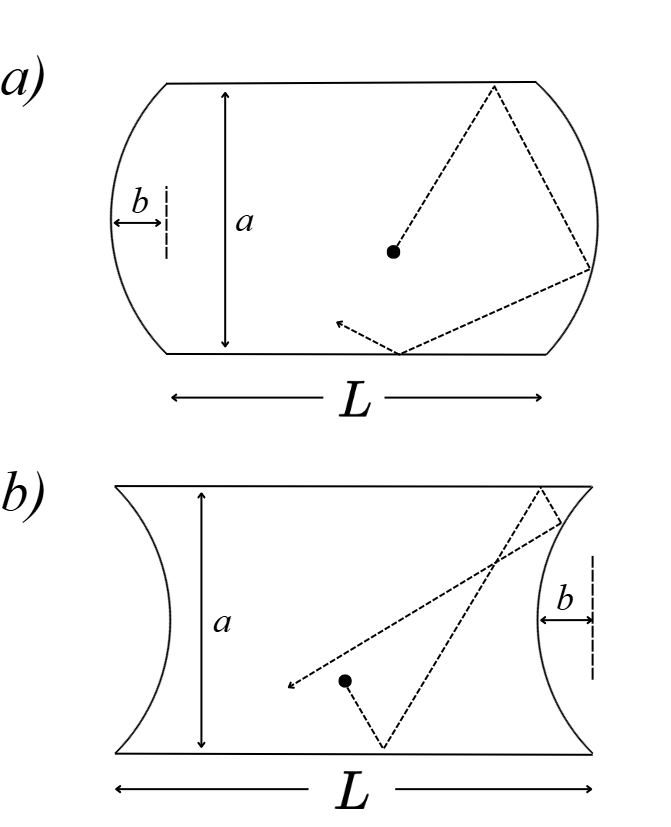}}
    \centerline{\includegraphics[width=0.65\linewidth]{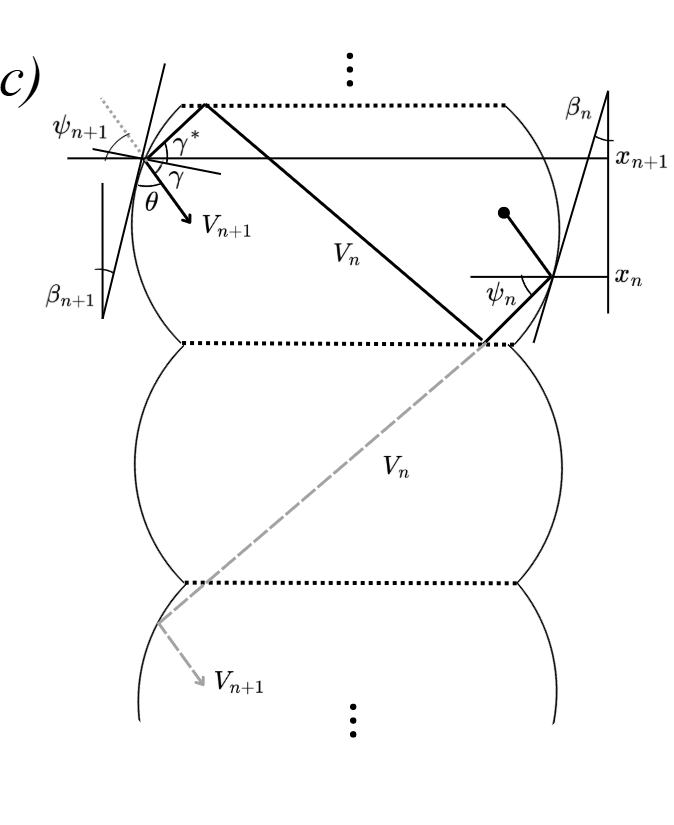}}
\caption{The stadium-like billiard and its geometric parameters with focusing (panel a) and defocusing (panel b) components. (c) Unfolding of the billiard presented in panel a and its dynamical variables}
\label{Fig2}
\end{figure}

The billiards considered in this work describe the motion of either a single particle or an ensemble of non-interacting particles confined within a closed boundary. For stadium-like geometries, the dynamics are described by a family of discrete maps. These systems traditionally consist of two focusing curves connected by two parallel straight segments (Fig.~\ref{Fig2}a) \cite{loskutov2002}. Here, we generalize this construction by also considering dispersing components (negative curvature), as illustrated in Fig.~\ref{Fig2}b.

Collisions with the straight segments in both billiard geometries correspond to simple specular reflections and do not contribute to the chaotic dynamics \cite{chernov2006chaotic}. To eliminate these dynamically trivial collisions, we employ the unfolding construction illustrated in Fig.~\ref{Fig2}c. Whenever the particle collides with a flat wall, the entire billiard table is reflected, allowing the trajectory to continue into its mirror image. Repeating this procedure after every collision with a straight segment effectively unfolds the billiard into an infinite sequence of mirrored cells. The resulting auxiliary trajectory collides with the curved boundaries at exactly the same angles as the original one, as shown in Fig.~\ref{Fig2}c, allowing the dynamics to be described exclusively by the focusing or dispersing components \cite{chernov2006chaotic}. The particle motion is therefore described by the mapping $T(x_n,\psi_n)=(x_{n+1},\psi_{n+1})$, where $x_n$ denotes the collision coordinate taken modulo $a$, and $\psi_n$ is the angle between the trajectory and the vertical direction, measured clockwise.

We assume that the curved boundaries are described by the quadratic function $f(x)=Ax^2+Bx+C$, where $A$, $B$, and $C$ are constants. The boundary conditions are identical for both the focusing and dispersing (defocusing) cases, namely $f(0)=0$ and $f(a)=0$, implying $C=0$ and $B=-Aa$. At the midpoint of the parabola, the conditions become $f(a/2)=-b$ for the focusing stadium (Fig.~2a) and $f(a/2)=b$ for the dispersing stadium (Fig.~2b). Consequently, $A=\pm 4b/a^2$ and $B=\mp 4b/a$ for panels (a) and (b), respectively. Substituting these coefficients into $f(x)$ yields the parabolic boundary
\begin{equation}
    \chi(x)=\pm \frac{4bx}{a^2}(x-a)
    \label{chi}
\end{equation}

To derive the mapping, we employ simple geometric arguments under the assumption that $l \gg b$, leading to
\begin{equation}
x_{n+1}=x_n+l \tan \psi_n
\end{equation}

To determine the angle $\psi_{n+1}$ after the collision, we define the auxiliary function $\beta(x)=\arctan[\chi'(x)]$. Expanding $\beta(x)$ to first order in a Taylor series yields
\begin{equation}
\beta(x) \approx \pm \frac{4b}{a^2}(2x-a)
\end{equation}

Introducing the dimensionless variable $\xi=x/a$, where $\xi \in [0,1)$, we recover the well-known map describing stadium-like billiards with static parabolic boundaries \cite{stadiumliv}:

\begin{equation}
    T:
\begin{cases}
\xi_{n+1}=\xi_n+\frac{l}{a}\tan \psi_n \mod 1\\[10pt]
\psi_{n+1}=\psi_n \mp\frac{8b}{a}(2\xi_{n+1}-1)
\end{cases}
\label{mapa}
\end{equation}
where the minus sign corresponds to the standard stadium-like billiard with focusing boundaries, whereas the plus sign corresponds to the dispersing case. The period-1 fixed points of the map $T$ are obtained by imposing $\xi_{n+1} = \xi_n + m$ and $\psi_{n+1} = \psi_n$, where $m \in \mathbb{Z}$ denotes the number of mirrored domains crossed by the particle in the unfolded representation. Solving these conditions yields $\xi^* = 1/2 + m$ and $\psi^* = \arctan(ma/l)$.

The stability of these fixed points is determined from the eigenvalues of the Jacobian matrix $J$ evaluated at $(\xi^*, \psi^*)$:
\begin{equation}
J =
\begin{pmatrix}
1 & \frac{l}{a \cos^2 \psi^*} \\[10pt]
\mp\frac{16b}{a} & 1 \mp \frac{16bl}{a^2 \cos^2 \psi^*}
\end{pmatrix}.
\end{equation}

Since $\det J = 1$, the mapping is area-preserving. The stability criterion for area-preserving maps states that a fixed point is elliptic, and therefore linearly stable, whenever the trace of the Jacobian matrix satisfies $|\text{Tr } J| \leq 2$ \cite{lichtenberg2013regular}. The trace is given by
\begin{equation}
\text{Tr } J = 2 \mp \frac{16bl}{a^2 \cos^2 \psi^*}.
\label{tr}
\end{equation}

For focusing boundaries, the trace becomes $\text{Tr } J = 2 - 16bl / (a^2 \cos^2 \psi^*)$. The fixed points lose stability when $\text{Tr } J < -2$. Considering the central fixed point, $m=0$ ($\psi^*=0$, $\cos^2 \psi^* = 1$), which is typically the last one to bifurcate, the condition for the destruction of all stable islands is \cite{stadiumliv}
\begin{equation}
\frac{4|b|l}{a^2} > 1.
\label{tr}
\end{equation}

Whenever this condition is satisfied, the phase space becomes globally chaotic. Conversely, if $4|b|l/a^2 \leq 1$, stable periodic islands coexist with the chaotic sea, giving rise to a mixed phase space. It is therefore convenient to define the critical value $|b_c|=a^2/4l$, at which the last regular structure disappears. For $|b|>|b_c|$, the phase space is fully chaotic.

\begin{figure}[h!]
  \centering
    \centerline{\includegraphics[width=\linewidth]{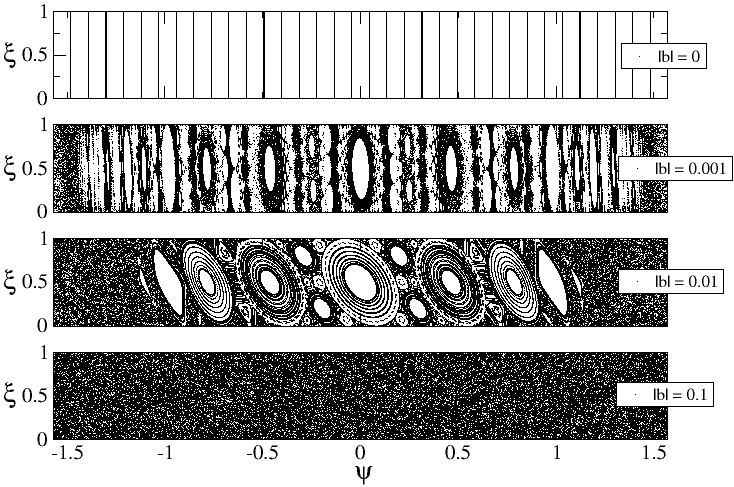}}
  \caption{Phase space for the static stadium-like billiard with focusing parabolic boundaries. The control parameters used were $a = 0.5$, $l = 1$ and different values of $b$ presented in each panel.} \label{Fig3}
\end{figure}

Phase-space portraits for the focusing boundaries are shown in Fig.~\ref{Fig3} for the control parameters $a=0.5$ and $l=1$, yielding $|b_c|=0.0625$. As expected, the phase space is fully integrable for $b=0$. As $|b|$ increases, chaotic regions emerge alongside invariant curves corresponding to regular motion. In the last panel, where $|b|=0.1 > |b_c|$, the phase space becomes fully chaotic, as predicted, marking the transition from local to global chaos.

Furthermore, Fig.~\ref{Fig3} reveals a dominant stable island centered at $\psi = 0$ (and $\xi = 1/2$). The robustness of this structure is supported by the existence of the primary period-1 fixed point ($m = 0$), which, according to the stability condition derived from Eq.~(\ref{tr}), is the last regular structure to lose stability before the onset of global chaos. It is also noteworthy that the chaotic sea is born globally connected. The absence of invariant spanning curves (global KAM tori) extending across the entire phase space can be explained by the singularities of the mapping \cite{katok1995introduction}. As shown by the Jacobian matrix $J$, the terms containing $1/\cos^2 \psi$ diverge as the particle approaches grazing collisions ($\psi \to \pm \pi/2$). In this limit, any potential spanning curves are destroyed and therefore cannot confine the chaotic motion to isolated regions. Consequently, the chaotic layer, initially more pronounced near $\psi \to \pm \pi/2$, remains unconstrained along the $\psi$ direction, permeating the entire phase space even for arbitrarily small values of $b>0$.

For the defocusing boundary, the trace becomes $\text{Tr } J = 2 + 16bl / (a^2 \cos^2 \psi^*)$. Since $b$, $l$, and $a$ are strictly positive, the second term is always positive. Therefore, $\text{Tr } J > 2$ for any $b > 0$, implying that the stability condition $|\text{Tr } J| \leq 2$ is never satisfied. Consequently, all fixed points are born hyperbolic, and the phase space is fully chaotic for any $b \neq 0$.

\begin{figure}[h]
  \centering
    \centerline{\includegraphics[width=\linewidth]{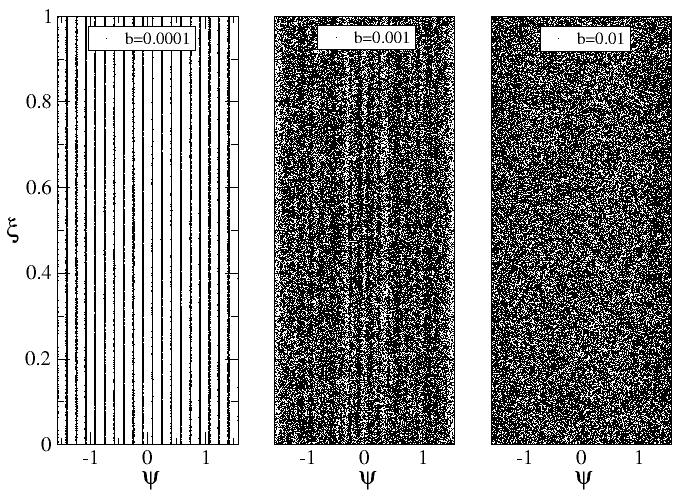}}
  \caption{Phase space for the static stadium-like billiard with defocusing parabolic boundaries. The control parameters used were $a = 0.5$, $l = 1$ and different values of $b$ presented in each panel.} \label{Fig4}
\end{figure}

Figure~\ref{Fig4} shows the corresponding phase-space portraits for the defocusing boundaries using the same control parameters, $a=0.5$ and $l=1$. As in the previous case, the phase space is completely integrable for $b=0$. However, for any $b>0$, the regular structures are immediately destroyed and the phase space becomes fully chaotic. This numerical result confirms the analytical prediction that stable fixed points cannot exist in this regime. For $b=0.0001$, remnants of the periodic structures are still visible in phase space. As $b$ increases, these remnants gradually disappear and the chaotic sea becomes increasingly homogeneous.

\section{Scaling properties and critical exponents}
\label{xsec3}

We now investigate the scaling behavior of the chaotic sea in the vicinity of the transition. To this end, we monitor the statistical evolution of the variable $\psi$. For a given initial condition $j$, the time average of $\psi$ after $n$ collisions is defined as
\begin{equation}
    \overline{\psi}_j(n,b) = \frac{1}{n}\sum^{n}_{i=1}\psi_{i,j},
\end{equation}
where the index $i$ denotes the $i$-th collision. This observable is particularly appropriate because $\psi$ plays the role of the momentum (or action) variable and is directly affected by the control parameter $b$, in close analogy with the kicked variable of the standard map \cite{leonel2020characterization}. By contrast, $\xi$ behaves as an angle variable and is topologically bounded by periodic boundary conditions, preventing unbounded diffusion.

To quantify this diffusion, we introduce the roughness $\omega$, defined as the interface width of the observable averaged over an ensemble of $M$ different initial conditions,
\begin{equation}
    \omega(n,b) = \frac{1}{M}\sum^{M}_{j=1}\sqrt{\overline{\psi^2}_j(n,b) -  \overline{\psi}_j^2(n,b)}.
    \label{omega}
\end{equation}

Another quantity of interest is the number of iterations required for the system to reach the stationary regime, corresponding to the time needed for trajectories to explore the entire accessible chaotic region of phase space. This characteristic time, commonly denoted by $n_x$, serves as an indirect measure of the correlation length near the transition. As the transition is approached, the divergence $n_x \rightarrow \infty$ indicates an increasing sensitivity to the initial conditions. Accordingly, the associated correlation length also diverges, a hallmark of continuous (second-order) phase transitions.

We now investigate the behavior of these observables across the two integrability-breaking transitions introduced in Section II: (i) from $b=0$ to $b<0$, corresponding to the introduction of focusing boundary components, and (ii) from $b=0$ to $b>0$, corresponding to the introduction of dispersing (defocusing) boundary components.

\subsection{Focusing component $(b<0)$}

\begin{figure}[h]
 \centering
   \centerline{\includegraphics[width=\linewidth]{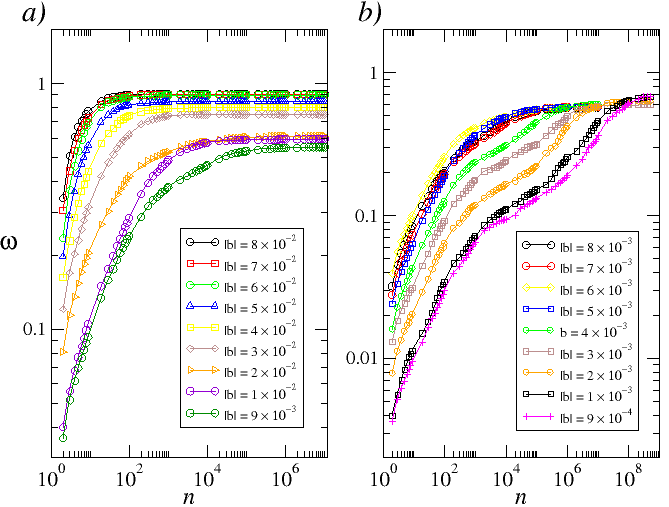}}
\caption{ Behavior of the roughness $\omega(n)$ as a function of the number of iterations $n$ for different values of the focusing control parameter. While scaling invariance is clearly observed in panel (a), panel (b) highlights the anomalous increase of the observable induced by stickiness.} \label{Fig5}
\end{figure}

Figure~\ref{Fig5} shows the behavior of the roughness $\omega(n)$ for different values of the control parameter $b$. The averages were computed over an ensemble of $10^3$ initial conditions uniformly distributed along a vertical strip $\xi_0 \in [0,1]$, with all trajectories initialized within the chaotic sea.

For $|b| > 9\times 10^{-3}$, as shown in Fig.~\ref{Fig5}(a), the expected scaling invariance is clearly observed. The roughness reaches the saturation value $\omega \approx 0.9068$ once $|b| > b_c = 0.0625$, where the accessible chaotic phase space has been completely explored. This limiting value is remarkably accurate, as it exactly coincides with the standard deviation of a uniform distribution over the domains $\xi \in [0,1]$ and $\psi \in [-\pi/2,\pi/2]$, namely $\pi/\sqrt{12}$.

\begin{figure}[h]
  \centering
    \centerline{\includegraphics[width=\linewidth]{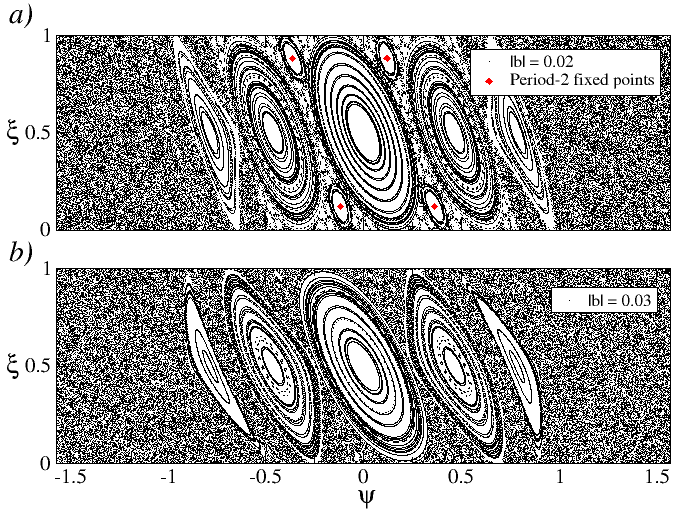}}
  \caption{Phase-space portraits highlighting the structural changes and the destruction of the period-2 islands for parameters corresponding to the gap observed in Fig.~\ref{Fig5}.}
  \label{Fig6nova}
\end{figure}

Figure~\ref{Fig5} also reveals a gap between the saturation plateaus for $|b| = 2\times 10^{-3}$ and $|b| = 3\times 10^{-3}$. This feature can be understood by analyzing the stability of the period-2 fixed points of the mapping. For a generic map $F$ satisfying $F(P_0)=P_1$ and $F(P_1)=P_0$, the determinant of the composed Jacobian matrix satisfies \cite{strog}:
\begin{equation}
    \det J_{2}(P) = \det J(P_1) \cdot \det J(P_0),
\end{equation}
leading to the following expression for the trace
\begin{equation}
     \text{Tr }J_2 = 2 - \frac{32|b|}{a}(A_0+A_1) + \frac{256|b|^2}{a^2}A_0 A_1,
\end{equation}
where $A_i = \frac{l}{a}(1+\tan^2\psi_i)$. Simultaneously, the coordinates of these fixed points can be obtained from the transcendental equation
\begin{equation}
     \arctan\left(\frac{a}{l}x\right) - \arctan\left(\frac{a}{l}(1-x)\right) = \frac{8|b|}{a}x,
\end{equation}
with $x = \xi_0 - \xi_1$ and $x \in [0,1]$. For the parameters considered here, the stability threshold $\text{Tr }J_2 = -2$ is reached at $|b| \approx 0.026$. Furthermore, imposing the maximum admissible value $x=1$ yields $|b| \approx 0.289$. Therefore, the gap observed in the roughness scaling reflects the change in stability and the subsequent destruction of the phase-space structures associated with these period-2 fixed points. This topological transition is further supported by the phase-space portraits shown in Fig.~\ref{Fig6nova}.

The behavior shown in Fig.~\ref{Fig5}(b) is dominated by stickiness, which induces the anomalous growth of $\omega(n)$. An orbit evolving in a chaotic region far from Kolmogorov-Arnold-Moser (KAM) islands explores phase space ergodically. In this regime, nearby initial conditions separate exponentially with time \cite{eckmann1985ergodic}. By contrast, stickiness occurs when a chaotic trajectory passes close to a KAM island and remains temporarily trapped, evolving almost regularly over exceptionally long time intervals \cite{Bunimovich_2008}. As can be seen in Fig.~\ref{Fig3}, the limit $|b| \rightarrow 0$ is accompanied by a proliferation of KAM islands and a progressive reduction of the chaotic sea near $\psi=\pm\pi/2$, thereby enhancing the influence of stickiness.

\begin{figure}[h]
  \centering
    \centerline{\includegraphics[width=\linewidth]{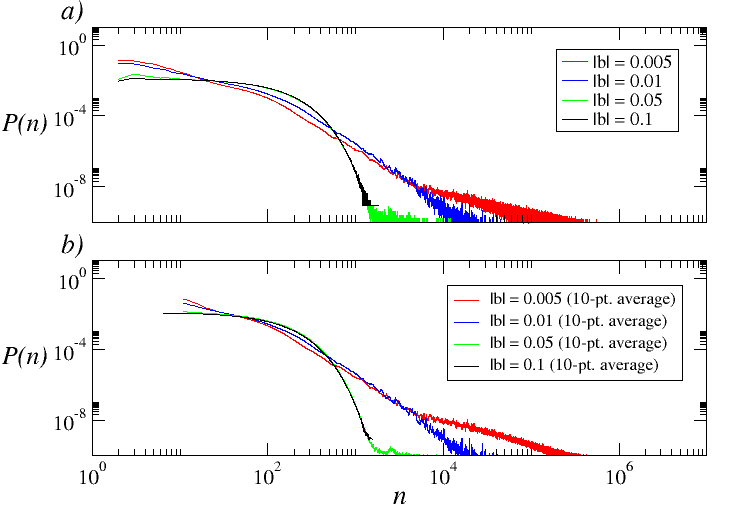}}
  \caption{Probability distribution $P(n)$ of the return times $n$ for different values of the control parameter. Panel (a) shows the raw normalized data, whereas panel (b) presents the corresponding 10-point moving average.}
  \label{Fig7nova}
\end{figure}

To quantify this enhancement, we investigate the probability distribution $P(n)$ of the return times. A single trajectory is initialized within the chaotic sea at the center of a box of dimensions $\Delta \xi = \Delta \psi = 10^{-1}$. We then measure the number of iterations $n$ required for the trajectory to return to this box, record the event, restart the counting procedure, and repeat the process over $10^{12}$ collisions. Normalizing the resulting histogram yields the probability distribution $P(n)$. The corresponding distributions are shown on a log-log scale in Fig.~\ref{Fig7nova} for different values of the control parameter. Panel (b) displays the same data after a 10-point moving average to reduce statistical fluctuations.

For a strongly chaotic and uniformly distributed sea ($|b|=0.1$), the expected exponential decay is observed. As $|b| \rightarrow 0$, an increasingly pronounced power-law tail emerges for large values of $n$, appearing as a linear regime on the log-log scale. This behavior confirms that trajectories become temporarily trapped in sticky regions of phase space, substantially increasing the probability of exceptionally long return times. Consequently, Fig.~\ref{Fig5}(b) demonstrates the significant influence of this anomalous transport mechanism on the roughness $\omega(n)$ for sufficiently small values of $|b|$. The localized trapping of trajectories strongly affects the ensemble averages entering Eq.~(\ref{omega}), giving rise to the anomalous growth observed in Fig.~\ref{Fig5} for $|b| < 9\times 10^{-3}$.

Returning to the behavior of $\omega(n)$ shown in Fig.~\ref{Fig5}, the curves initially exhibit a power-law growth governed by chaotic diffusion for short iteration times ($n \ll n_x$, where $n_x$ denotes the crossover time). After a sufficiently large number of collisions, all curves undergo a crossover toward a saturation plateau, eventually reaching a stationary regime. The saturation values satisfy the scaling relation $\omega_{sat} \propto |b|^{\alpha_1}$, where $\alpha_1$ denotes the saturation exponent.

\begin{figure}[h]
  \centering
    \centerline{\includegraphics[width=\linewidth]{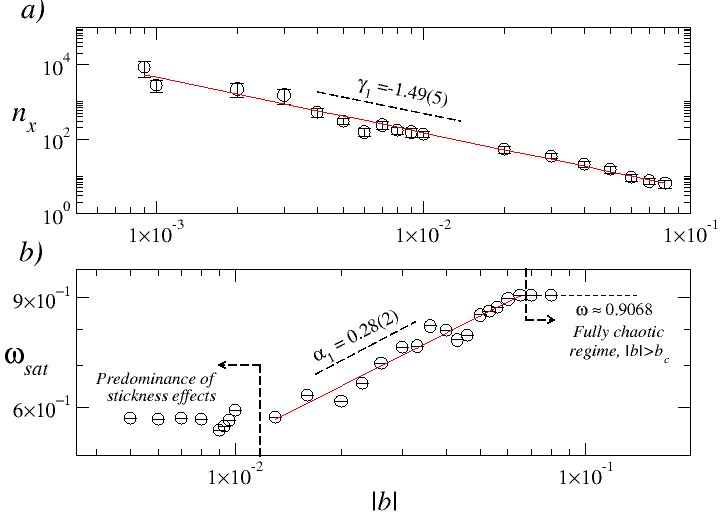}}
  \caption{Behavior of the observables $n_x$ and $\omega_{sat}$ as a function of the control parameter $|b|$ for the focusing case, associated power-law fit and exponents $\gamma_1=1.49(5)$ and $\alpha_1=0.28(2)$, respectively.}
  \label{Fig8}
\end{figure}

The crossover time $n_x$ is determined by fitting the first decade of points of each curve with a power law. We then define $n_x$ as the intersection between this initial growth law and the corresponding saturation plateau $\omega_{sat}$. Figure~\ref{Fig8}(a) presents the dependence of $n_x$ on $|b|$. The resulting power-law fit yields the exponent $\gamma_1 = -1.49(5)$, confirming the expected divergence of $n_x$ as the integrability limit ($b=0$) is approached. As discussed above, this divergence provides an indirect measure of the correlation length and constitutes a characteristic signature of a continuous phase transition.

We also extract the asymptotic saturation values $\omega_{sat}$ for different values of the control parameter, as shown in Fig.~\ref{Fig8}(b). As discussed previously, the dynamics for small values of $|b|$ are strongly affected by stickiness, which artificially enhances the roughness. Conversely, for $|b| > b_c$, the system becomes fully chaotic and $\omega_{sat}$ reaches the theoretical plateau $\pi/\sqrt{12} \approx 0.9068$. By fitting the intermediate scaling regime between these two limits, we obtain the critical saturation exponent $\alpha_1 = 0.28(2)$.

This result reveals a distinct universality class for the transition from integrability to non-integrability, differing from that observed in systems such as the Standard Map and the Fermi-Ulam model \cite{bridge}. In those systems, the nonlinear perturbation generates resonance islands whose characteristic width scales with the square root of the perturbation parameter (i.e., $\alpha = 0.5$). The present system, however, exhibits a fundamentally different phase-space topology. Here, the chaotic sea does not emerge from the destruction of a separatrix that develops into a stochastic layer \cite{denisnovo}. Instead, it is born globally connected, occupying the available phase-space region between the KAM islands that appear as soon as $b \neq 0$. Moreover, in systems governed by anomalous transport, the evolution of the effective coarse-grained phase-space volume is strongly influenced by complex trapping structures \cite{ZASLAVSKY2002461}. As a consequence, the statistical recurrence responsible for the growth of the explored volume is modified, leading to non-integer scaling exponents. Since the roughness $\omega$ measures the geometric dispersion of the phase-space region explored by the trajectories, the measured critical exponent suggests that the integrability-breaking transition belongs to an anomalous universality class governed by fractional transport dynamics.

\subsection{Defocusing component $(b>0)$}

\begin{figure}[h]
  \centering
    \centerline{\includegraphics[width=\linewidth]{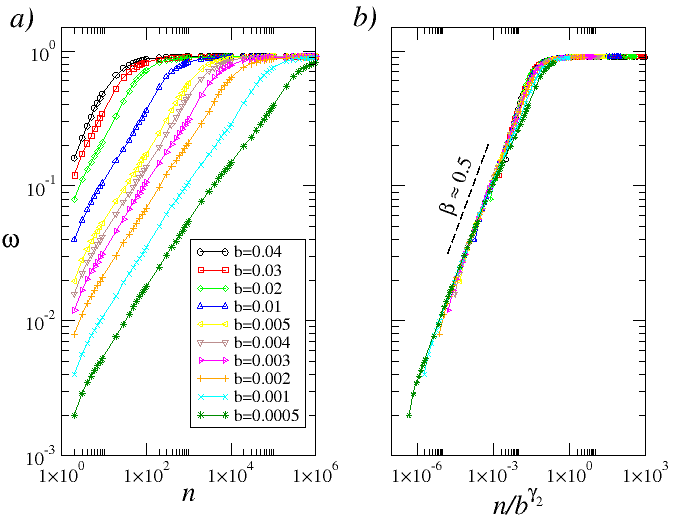}}
  \caption{Behavior of the roughness $\omega(n)$ as a function of the number of iterations $n$ for different values of the defocusing control parameter (a), and the corresponding collapse of the curves after the scaling transformation $n \rightarrow n/b^{\gamma_2}$ (b).}
  \label{Fig9}
\end{figure}

For the dispersing (defocusing) boundary components, the behavior of $\omega(n)$ is shown in Fig.~\ref{Fig9}. Unlike the focusing case, the curves exhibit an initial power-law growth for short iteration times ($n \ll n_x$), governed by normal chaotic diffusion according to $\omega(n) \propto n^\beta$, with $\beta \approx 0.5$. For sufficiently long iteration times ($n \gg n_x$), all curves bend toward a common saturation plateau, reaching the theoretical value $\omega_{\text{sat}} \approx 0.9068$. Consequently, the saturation exponent vanishes, yielding $\alpha_2 \approx 0$.

This vanishing exponent reflects a fundamental physical distinction from the focusing geometry. Since the phase space is fully chaotic for any $b \neq 0$, it contains no regular structures such as KAM islands. Consequently, chaotic trajectories diffuse uniformly throughout the entire accessible phase space, and the saturation width is determined solely by the system's geometric boundaries rather than by the perturbation parameter $b$. Nevertheless, $b$ continues to play an important dynamical role by determining the number of iterations required to reach the stationary state. Increasing $b$ enhances the curvature of the dispersing boundaries, thereby strengthening the scattering mechanism, accelerating diffusion, and reducing the crossover time $n_x$. This dependence is described by the scaling relation $n_x \propto b^{\gamma_2}$, where $\gamma_2$ denotes the crossover exponent.

To describe the observed behavior of $\omega(n)$ in the defocusing billiard analytically, we model the dynamics as a stochastic process governed by a one-dimensional diffusion equation. Let $P(\psi, n)$ denote the probability density of finding a particle with angular coordinate $\psi$ after the $n$-th collision. Since the phase space is bounded by $\psi \in [-L, L]$, where $L=\pi/2$, and assuming reflecting boundary conditions, the exact solution of the diffusion equation for an ensemble initially concentrated around $\psi \approx 0$ can be written as the Fourier series expansion \cite{butkov}
\begin{equation}
    P(\psi, n) = \frac{1}{2L} + \frac{1}{L} \sum_{k=1}^{\infty} \cos\left(\frac{k\pi\psi}{L}\right) \exp\left( - \frac{k^2 \pi^2 D n}{L^2} \right),
    \label{distrib}
\end{equation}
where $D$ denotes the diffusion coefficient. As will be shown in Sec.~\ref{xsec4}, the characteristic step size of the random walk scales linearly with the geometric perturbation parameter $b$, implying that $D \propto b^2$.

The scaling behavior of $\omega(n,b)$ follows directly from Eq.~(\ref{distrib}). For short iteration times ($n \ll n_x$), particles diffuse freely before reaching the physical boundaries of phase space. In this limit, the Fourier series converges to a Gaussian distribution, yielding normal diffusion with $\omega(n,b) \propto b n^\beta$. Since $\omega = \sqrt{\langle \psi^2 \rangle} \propto \sqrt{Dn}$ and $D \propto b^2$, it immediately follows that $\omega \propto b n^{1/2}$ and therefore $\beta=1/2$, in excellent agreement with the numerical results.

For sufficiently long time scales ($n \gg n_x$), the system reaches a stationary state. Mathematically, as $n \rightarrow \infty$, all time-dependent exponential terms in Eq.~(\ref{distrib}) vanish, and the probability density converges to the uniform distribution $P(\psi,n)_{n\rightarrow\infty}=1/2L$. The corresponding saturation value can then be obtained directly, $\omega_{\text{sat}} = \sqrt{\int_{-L}^{L} \psi^2 (2L)^{-1} d\psi} = L/\sqrt{3}$.

Substituting $L=\pi/2$ yields the analytical result $\omega_{\text{sat}}=\pi/\sqrt{12}\approx0.9068$. Importantly, this saturation value depends exclusively on the geometric extent of the accessible phase space and is completely independent of the perturbation parameter $b$. Consequently, the saturation exponent vanishes in this geometry ($\alpha_2=0$), leading to $\omega_{\text{sat}} \propto b^0=\text{constant}$.

The crossover from the initial power-law growth to the saturation regime occurs at the characteristic time $n_x$. At this point, $\omega(n_x,b)\sim\omega_{\text{sat}}$, or equivalently, $b n_x^\beta \propto \text{constant}$. Since the crossover time scales as $n_x \propto b^{\gamma_2}$, the relation between the critical exponents becomes
\begin{equation}
    b \left( b^{\gamma_2} \right)^\beta \propto b^0 \implies b^{1 + \gamma_2 \beta} \propto b^0.
\end{equation}

This immediately leads to the analytical scaling law
\begin{equation}
    \gamma_2 = -\frac{1}{\beta}.
    \label{leideescala}
\end{equation}

Our independent numerical simulations for the defocusing geometry yield the growth exponent $\beta=0.5$. Substituting this value into Eq.~(\ref{leideescala}) predicts the crossover exponent $\gamma_2=-1/0.5\approx-2$. Moreover, applying the scaling transformation $n \rightarrow n/b^{\gamma_2}$ produces an excellent collapse of all curves onto a single universal scaling function, as shown in Fig.~\ref{Fig9}(b). This agreement provides strong support for the statistical-mechanics description of the integrability-breaking transition in this geometry.

In summary, we have characterized the scaling behavior associated with the integrability-breaking transition ($b=0 \rightarrow b\neq0$) for both focusing and dispersing stadium-like billiards. By analyzing the evolution of the roughness $\omega$ within the chaotic sea, we obtained the critical exponent $\alpha_1=0.28(2)$ for the focusing geometry, where the dynamics are strongly influenced by stickiness and fractional transport. In contrast, for the strictly ergodic dispersing geometry, we obtained $\alpha_2\approx0$ together with the crossover exponent $\gamma_2=-1/\beta=-2$. For this latter case, these critical exponents enable all numerical data to collapse onto a single universal scaling curve. Such scaling invariance, together with the associated critical exponents, constitutes a hallmark of continuous phase transitions, a connection that will be explored in the next section.

\section{The phase transitions}
\label{xsec4}

Two distinct phase transitions are observed in the systems investigated here. For the stadium-like billiard with focusing parabolic boundaries, the transition is continuous as $b \rightarrow 0$. As the control parameter increases, the system evolves from an integrable regime to a mixed phase space characterized by the progressive growth of the chaotic sea and the gradual destruction of regular structures. Once $|b|>b_c$, as discussed in Sec.~\ref{xsec2}, the phase space eventually becomes fully chaotic. To characterize this continuous transition, we address four fundamental questions commonly employed in statistical mechanics \cite{sethna2021statistical}: (1) What symmetry is broken at the transition? (2) What is the appropriate order parameter, and does its susceptibility diverge at criticality? (3) What are the elementary excitations responsible for particle diffusion? and (4) What are the topological defects governing transport across phase space?

To address the first question, we refer to the phase-space portraits shown in Fig.~\ref{Fig3}. In the integrable limit ($b=0$), the phase space is foliated by continuous invariant curves, reflecting the continuous symmetry of the unperturbed invariant tori. Introducing the control parameter ($b \neq 0$) breaks this symmetry and drives the system into a mixed regime. The invariant curves are progressively destroyed and replaced by periodic islands embedded in a chaotic sea. Therefore, the broken symmetry associated with this transition is the continuous topological symmetry of the invariant foliation, which is destroyed by the geometric perturbation as the system evolves from integrability to non-integrability.

The next step is to identify an order parameter capable of distinguishing the two dynamical phases. Since the transition is governed by diffusive transport, a natural choice is the saturation roughness $\omega_{\text{sat}}$, which obeys the scaling relation $\omega_{\text{sat}} \propto b^{\alpha_1}$ close to the transition. As shown in Sec.~\ref{xsec3} (Fig.~\ref{Fig8}), the corresponding power-law fit yields $\alpha_1=0.28(2)$, demonstrating that the order parameter vanishes continuously as $b\rightarrow0$. The corresponding susceptibility is
\begin{equation}
    \chi = \frac{d\omega_{\text{sat}}}{db} \propto \frac{\alpha}{b^{1-\alpha}}.
\end{equation}

Since $0<\alpha_1<1$, the exponent $(1-\alpha_1)$ is positive, implying that $\chi$ diverges as $|b|\rightarrow0$. Furthermore, the crossover time $n_x$ plays the role of an effective correlation length for this transition. Its divergence, $n_x\propto|b|^{\gamma_1}$ with $\gamma_1<0$, is directly analogous to the critical slowing down observed in conventional critical phenomena. The simultaneous occurrence of a continuously vanishing order parameter, a diverging susceptibility, and a diverging correlation time provides strong quantitative evidence that the integrability-breaking process constitutes a genuine continuous phase transition.

We now turn to the elementary excitations responsible for chaotic diffusion. Inspection of the mapping shows that the control parameter $b$ determines the amplitude of the nonlinear perturbation. Diffusion is triggered as $\psi\rightarrow\pi/2$, where the term $\tan\psi_n$ in the first equation diverges. As a consequence, the variable $\xi_{n+1}$ wraps around the unit interval ($\mod1$) at an extremely high rate, effectively destroying any sequential correlation in $\xi$. Consequently, the quantity $(2\xi_{n+1}-1)$ appearing in the second equation behaves as an uncorrelated random variable uniformly distributed over the interval $[-1,1]$. Under these conditions, the dynamics of $\psi$ become equivalent to a random-walk process, producing chaotic diffusion throughout phase space. Moving away from this completely uncorrelated limit, correlations are gradually restored and regular structures, such as periodic islands and invariant curves, re-emerge.

The characteristic amplitude of these excitations can be estimated from the root-mean-square of the second equation of the mapping in Eq.~(\ref{mapa})
\begin{equation}
    \overline{\psi^2_{n+1}}= \overline{\psi^2_{n}}+ \left(\frac{8b}{a}\right)^2 \overline{(2\xi_{n+1}-1)^2}.
\end{equation}

Assuming that the ensemble is initially prepared close to the unperturbed boundary, such that $\overline{\psi_n^2}\approx0$, we obtain
\begin{equation}
    \psi_a=\frac{8b}{a\sqrt{3}},
\end{equation}
since $\overline{(2\xi_{n+1}-1)^2}=1/3$. Therefore, the elementary excitations of the system are identified with the geometric perturbation term $-\frac{8b}{a}(2\xi_{n+1}-1)$ appearing in the second equation of the mapping, whose characteristic amplitude is given by $\psi_a$. This perturbation acts as the microscopic mechanism responsible for generating the chaotic diffusion of $\psi$.

Finally, we identify the topological defects governing transport across phase space. In the present system, these defects correspond to the periodic islands. Their influence on transport arises through the stickiness phenomenon, whereby chaotic trajectories become temporarily trapped near the boundaries of regular islands, leading to a local enhancement of trajectory density. These islands are centered on elliptic fixed points. As established previously, the trace of the Jacobian matrix is given by $\text{Tr } J = 2 - 16bl / (a^2 \cos^2 \psi^*)$. These stable fixed points appear for $b\neq0$ and subsequently lose stability once $\text{Tr } J<-2$, corresponding to $b_c=a^2/4l$. Beyond this threshold, the last regular structures disappear and the phase space becomes completely chaotic. Periodic islands therefore constitute the topological defects that govern particle transport prior to the onset of global chaos.

By contrast, the stadium-like billiard with dispersing boundary components undergoes a first-order transition occurring abruptly at $b=0$. This discontinuous change is clearly illustrated by the phase-space portraits shown in Fig.~\ref{Fig4}. The fully integrable rectangular billiard immediately gives way to a globally chaotic phase space for any $b\neq0$. Through the unfolding construction, the billiard may be interpreted as a finite-horizon semi-dispersing system with corner singularities. Since the boundary is globally dispersing, the dynamics belong to the class of planar semi-dispersing billiards with singularities, which are rigorously known to be hyperbolic and ergodic \cite{ChernovSimanyi2010}. This mathematical characterization is fully consistent with the scaling analysis presented in Sec.~\ref{xsec3}, where the analytical relation $\gamma_2=-1/\beta$, together with $\beta=0.5$, yields $\gamma_2=-2$, reflecting the normal diffusion expected for a fully chaotic phase space. The excellent agreement between the analytical prediction and the numerical results confirms that the dispersing billiard is globally chaotic and devoid of visible regular structures.

\section{Summary and conclusions}
\label{xsec5}

We have investigated the transition from integrability to non-integrability in two classes of stadium-like billiards distinguished by the curvature of their parabolic boundaries. Our results reveal two qualitatively different dynamical transitions. For focusing boundaries, the system undergoes a continuous transition from an integrable phase to a mixed phase space in which regular KAM islands coexist with a chaotic sea, eventually evolving into global chaos for $|b|>|b_c|$. In contrast, dispersing boundaries produce an abrupt transition: the integrable rectangular billiard immediately becomes fully chaotic for any finite perturbation ($b\neq0$), characterizing a first-order transition.

The scaling properties of the chaotic dynamics were characterized through the evolution of the roughness $\omega(n)$. For the dispersing geometry, we obtained a vanishing saturation exponent, $\alpha_2\approx0$, together with the crossover exponent $\gamma_2=-2$, derived analytically from the scaling law $\gamma_2=-1/\beta$. These exponents successfully collapse all numerical results onto a single universal scaling curve. For the focusing geometry, we identified an anomalous saturation exponent, $\alpha_1=0.28(2)$, associated with the divergence of the crossover time and the emergence of anomalous transport induced by stickiness.

Beyond the scaling analysis, we established a complete physical characterization of the continuous integrability-breaking transition by addressing the four fundamental ingredients of critical phenomena. We identified the broken symmetry as the destruction of the continuous invariant foliation of the integrable phase space. The saturation roughness, $\omega_{\text{sat}}\propto|b|^{\alpha_1}$, was shown to constitute an appropriate order parameter whose susceptibility diverges at criticality, while the crossover time $n_x$ plays the role of an effective correlation length through its critical divergence. We further identified the microscopic elementary excitations responsible for chaotic diffusion as the geometric momentum kicks generated by the perturbation term $-(8b/a)(2\xi_{n+1}-1)$ and demonstrated that the periodic islands act as the relevant topological defects by inducing stickiness and anomalous transport throughout the mixed phase space.

These results provide a unified statistical-mechanics framework for describing integrability-breaking transitions in Hamiltonian billiards. More broadly, they suggest that concepts traditionally associated with equilibrium critical phenomena—including broken symmetry, order parameters, elementary excitations, topological defects, and critical scaling—can also be naturally extended to deterministic nonlinear dynamical systems. We expect that this perspective may prove useful for investigating analogous transitions in a broad class of Hamiltonian systems beyond billiard dynamics.\\

\section*{Acknowledgements}
A.K.P.F. acknowledges CAPES (No. $88887.990665/2024-00$) for financial support. E.D.L. acknowledges support from Brazilian agencies CNPq (No. $301318/2019-0, 304398/2023-3)$ and FAPESP (No. $2019/14038-6$ and No. $2021/09519-5)$.

\bibliography{references}% Produces the bibliography via BibTeX.

\end{document}